\documentclass[prl,aps,floatfix,showpacs,twocolumn]{revtex4-1}
\usepackage{dcolumn}
\usepackage{bm}
\usepackage{color}
\usepackage{graphicx}
\usepackage{epsf}
\usepackage{pstricks}
\begin{document}

\title{Thermal neutron captures on $d$ and $^3$He}
\author{L.\ Girlanda$^{\,{\rm a,b}}$, A.\ Kievsky$^{\,{\rm b}}$, L.E.\ Marcucci$^{\, {\rm a,b}}$,
S.\ Pastore$^{\,{\rm c}}$, R.\ Schiavilla$^{\,{\rm c,d}}$, and M.\ Viviani$^{\,{\rm b}}$}
\affiliation{
$^{\,{\rm a}}$\mbox{Department of Physics, University of Pisa, 56127 Pisa, Italy}\\
$^{\,{\rm b}}$\mbox{INFN-Pisa, 56127 Pisa, Italy}\\
$^{\,{\rm c}}$\mbox{Department of Physics, Old Dominion University, Norfolk, VA 23529, USA}\\
$^{\,{\rm d}}$\mbox{Jefferson Lab, Newport News, VA 23606}\\
}

\date{\today}

\begin{abstract}
We report on a study of the $nd$ and $n\,^3$He radiative captures at thermal
neutron energies, using wave functions obtained from either chiral or conventional
two- and three-nucleon realistic potentials with the hyperspherical harmonics method,
and electromagnetic currents derived in chiral effective field theory
up to one loop.  The predicted $nd$ and $n\,^3$He cross sections
are in good agreement with data, but exhibit a significant dependence on the input
Hamiltonian.  A comparison is also made between these and new results for the
$nd$ and $n\,^3$He cross sections obtained in the conventional framework for
both potentials and currents.
\vspace{-.5cm}
\end{abstract}


\index{}\maketitle
The $nd$ and $n\,^3$He radiative capture reactions at thermal
neutron energy are very interesting, in that the magnetic dipole ($M1$)
transitions connecting the continuum states to the hydrogen and helium
bound states are inhibited at the one-body level.  Hence, most of the
calculated cross sections (80--90\% in the case of $n\, ^3$He) results
from contributions of many-body components in the electromagnetic
current operator~\cite{Marcucci05}.  Thus these processes provide a
crucial testing ground for models describing these many-body
operators and, indirectly, the nuclear potentials from which the ground-
and scattering-state wave functions are derived.

Over the past two decades, chiral effective field theory ($\chi$EFT), originally
proposed by Weinberg in a series of papers in the early nineties~\cite{Weinberg90},
has blossomed into a very active field of research.  The chiral symmetry exhibited by
quantum chromodynamics (QCD) severely restricts the form of the interactions
of pions among themselves and with other particles.  In particular,
the pion couples to baryons, such as nucleons and $\Delta$-isobars,
by powers of its momentum $Q$, and the Lagrangian describing
these interactions can be expanded in powers of $Q/\Lambda_\chi$, where
$\Lambda_\chi \sim 1$ GeV specifies the chiral-symmetry breaking
scale.  As a result, classes of Lagrangians emerge, each characterized
by a given power of $Q/\Lambda_\chi$ and each involving a certain number of
unknown coefficients, so called low-energy constants (LEC's), which are then
determined by fits to experimental data (see, for example, the
review papers~\cite{Bedaque02} and~\cite{Epelbaum09}, and
references therein).  Thus, $\chi$EFT provides, on the one hand,
a direct connection between QCD and its symmetries, in particular
chiral symmetry, and the strong and electroweak interactions in nuclei,
and, on the other hand, a practical calculational scheme susceptible,
in principle, of systematic improvement.  In this sense, it can
be justifiably argued to have put low-energy few-nucleon physics
on a more fundamental basis.

Concurrent with these conceptual developments have been the acquisition and
refinement of accurate methods for solving the $A$=3 and 4 Schr\"odinger equation
(see Ref.~\cite{Carlson98} for a review).  In this respect, it is worthwhile
noting that the $A$=4 scattering problem has proven to be especially
challenging for two reasons.  The first is its coupled-channel nature: even
at vanishing energies for the incident neutron, the elastic $n$-$^3$He and
charge-exchange $p$-$^3$H channels are both open, and need to be
accounted for.  The second reason lies in the peculiarities of the $^4$He spectrum,
in particular the presence of resonant states between the $p$-$^3$H and
$n$-$^3$He thresholds, which make it hard to obtain numerically converged
solutions.  Indeed, it is only very recently that both these capabilities have been
fully realized~\cite{Deltuva07,Viviani10}. In the present work, the 3- and
4-body problems are solved with the hyperspherical-harmonics (HH) technique
(see Ref.~\cite{Kievsky08} for a review).

The developments outlined above make it possible to examine the question
of whether available experimental data on these delicate processes---the $nd$ and
$n\,^3$He captures---are well reproduced by theory.  The present letter reports
on such an effort by presenting results obtained both in $\chi$EFT as well
as in the conventional framework based (essentially) on a meson-exchange
model of potentials and electromagnetic current operators.  This approach,
while more phenomenological than $\chi$EFT, has a broader range of applicability,
and accounts satisfactorily for a wide variety of
nuclear properties and reactions up to energies, in some cases, beyond
the pion production threshold, see Ref.~\cite{Carlson98} for a review.
In particular, it reproduces well observed magnetic properties
of $A=2$ and 3 nuclei, including moments and form factors, as well as
the $np$ radiative capture, see Marcucci {\it et al.} (2005)~\cite{Marcucci05}.

The model for the nuclear electromagnetic current in $\chi$EFT up to one loop was derived
originally by Park {\it et al.}~\cite{Park96}, using covariant perturbation
theory.  In the last couple of years, two independent derivations, based
on time-ordered perturbation theory (TOPT), have appeared in the literature, one
by K\"olling {\it et al.}~\cite{Koelling09} and the other by some of the
present authors~\cite{Pastore09}.  There are technical differences in the
implementation of TOPT, which relate to the treatment of reducible diagrams
and are documented in considerable detail in the above papers.
However, the resulting expressions in Refs.~\cite{Koelling09} and~\cite{Pastore09}
for the two-pion-exchange currents (the only ones considered by the authors
of Ref.~\cite{Koelling09}) are in agreement with each other, but differ
from those of Ref.~\cite{Park96}, in particular in the isospin structure
of the $M1$ operator associated with the one-loop corrections---see
Pastore {\it et al.} (2009)~\cite{Pastore09} for a comparison
and analysis of these differences.

Explicit expressions for the $\chi$EFT currents up to one loop, and associated $M1$ operators,
are listed in Refs.~\cite{Pastore09}.  Here we summarize succinctly their main features.
The leading-order (LO) term results from the coupling of the external photon field to the individual nucleons,
and is counted as $e\,Q^{-2}$ ($e$ is the electric charge).  The NLO term
(of order $e\, Q^{-1}$) involves seagull and in-flight
contributions associated with one-pion exchange, and the N$^2$LO term (of order $e\, Q^0$)
represents the $(Q/m_N)^2$ relativistic correction to the LO one-body current ($m_N$ denotes the
nucleon mass).
\vspace{-1cm}
\begin{widetext}
\vspace{-0.5cm}
\begin{center}
\begin{figure}[bthp]
\includegraphics[width=5.5in]{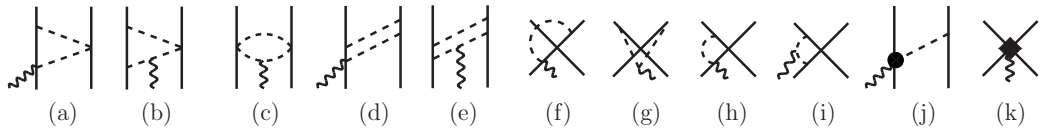}
\vspace{-0.4cm}
\caption{Diagrams illustrating two-body currents at N$^3$LO.  Nucleons,
pions, and photons are denoted by solid, dashed, and wavy lines, respectively.
Only one among the possible time orderings is shown for diagrams (a)-(j).}
\label{fig:f1}
\end{figure}
\end{center}
\vspace{-1cm}
\end{widetext}
\vspace{-1cm}

At N$^3$LO ($e\, Q$) we distinguish three classes of terms~\cite{Pastore09}:
i) two-pion exchange currents at one loop, illustrated by diagrams (a)-(i) in
Fig.~\ref{fig:f1}, ii) a tree-level one-pion exchange current involving the
standard $\pi NN$ vertex on one nucleon, and a $\gamma \pi NN$ vertex of order
$e\, Q^2$ on the other nucleon, illustrated by diagram (j), and iii) currents
generated by minimal substitution in the four-nucleon contact interactions
involving two gradients of the nucleons' fields as well as by non-minimal 
couplings, collectively represented by diagram (k).  A fourth class consisting of $(Q/m_N)^2$ relativistic
corrections (RC's) to the NLO currents is neglected.  However, RC's are not consistently treated
in available chiral potentials, such as those employed below.  For example,
the RC's in the two-nucleon potential, implied by Poincar\'e covariance
and just derived in Ref.~\cite{Girlanda10} in an EFT context, have
been omitted so far in $A$=3 and 4 calculations, even though their
contribution is expected to be comparable to that of the three-nucleon potential~\cite{Girlanda10}.

The loop corrections in panels (a)-(i) involve the pion mass, the nucleon axial
coupling constant $g_A$=1.29 (from the Goldberger-Treiman relation
relating it to the $\pi NN$ coupling constant), and the pion decay amplitude
$F_\pi$=184.8 MeV.  The LEC's entering panel (i) and the minimal currents
in panel (k) have been determined by fits to the $np$ S- and P-wave phase
shifts up to 100 MeV laboratory energies~\cite{Pastore09}.  We refer below
to these constrained terms collectively as  N$^3$LO(S-L).  There are five
additional unknown LEC's: $d_8^{\, \prime}$, $d_9^{\, \prime}$, and
$d_{21}^{\, \prime}$ in panel (j), and $C_{15}^\prime$ and $C_{16}^\prime$
in the non-minimal currents of panel (k).  We denote these terms as N$^3$LO(LECs)
in the following.  In a resonance saturation picture, the
$d_8^{\, \prime}$ and $d_{21}^{\, \prime}$ ($d_9^{\, \prime}$) LEC's
can be related to the combination of coupling constants and masses entering
the isovector (isoscalar) $N$-$\Delta$ excitation and $\omega\pi\gamma$ ($\rho\pi\gamma$)
transition currents~\cite{Pastore09}.  Indeed, this connection is exploited in a series of calculations,
based on the $M1$ operators derived in Ref.~\cite{Park96}, of the $np$, $nd$, and $n\, ^3$He radiative
captures, and magnetic moments of $A$=2 and 3 nuclei~\cite{Park96,Park00}.
Here, however, we adopt a different strategy, as discussed below.
Lastly, we observe that at N$^3$LO there are no three-body currents
in the formalism of Ref.~\cite{Pastore09}, which retains irreducible and
recoil-corrected reducible diagrams.

The $\chi$EFT $M1$ operators have power-law behavior
for large relative momenta $k$'s, and need to be
regularized, before they can be inserted between nuclear wave functions.
Following common practice, we implement this regularization
by means of a cutoff $C_\Lambda(k)={\rm exp}(-k^4/\Lambda^4)$,
with $\Lambda$ in the range (500--700) MeV, and constrain
the LEC's entering the N$^3$LO $M1$ operators of panels (j) and (k)
in Fig.~\ref{fig:f1} to reproduce a set of observables for any
given $\Lambda$ in this range.  This same renormalization procedure
is adopted in the currently most advanced analyses of nuclear
potentials, for example in Ref.~\cite{Entem03}
(see also Sec.~II.C of Ref.~\cite{Epelbaum09}, and references therein,
for further discussion of this issue).

These operators are used in the present work
to study the magnetic moments of the deuteron and trinucleons,
and the $np$, $nd$, and $n\,^3$He radiative captures  at thermal neutron
energies.  The calculations are carried out by evaluating their matrix elements
between wave functions obtained from either conventional or chiral
(realistic) potentials with the variational HH method~\cite{Kievsky08}.
We consider the Argonne $v_{18}$~\cite{Wiringa95} (AV18) and chiral N$^3$LO~\cite{Entem03}
(N3LO) two-nucleon potentials in combination with the Urbana-IX~\cite{Pudliner97}
(UIX) and chiral N$^2$LO~\cite{Gazit09} (N2LO) three-nucleon potentials.
The AV18/UIX and N3LO/N2LO Hamiltonians provide a very good description
of three- and four-nucleon bound and scattering state properties, including
binding energies, radii, and scattering lengths~\cite{Viviani10,Kievsky08}.

We now turn our attention to the determination of the LEC's $d_8^{\, \prime}$,
$d_9^{\, \prime}$, $d_{21}^{\, \prime}$,  $C_{15}^\prime$, and $C_{16}^\prime$.
In principle, the $d_i^{\, \prime}$ could be fitted to pion photoproduction
data on a single nucleon or, as mentioned already, related to hadronic
coupling constants by resonance saturation arguments.
Both procedures have drawbacks.  While the former achieves consistency with the
single-nucleon sector, it nevertheless relies on single-nucleon
data involving photon energies much higher than those relevant to the threshold processes under
consideration and real (in contrast to virtual) pions (some of these
same issues in the context of three-nucleon potentials have been
investigated in Ref.~\cite{Pandharipande05}).  The second procedure is
questionable because of poor knowledge of some of the
hadronic couplings, such as $g_{\omega NN}$ and $g_{\rho NN}$.

Here, we assume $d_{21}^{\, \prime}/d_8^{\, \prime}=1/4$ as suggested
by $\Delta$-dominance, and rely on nuclear data to constrain the remaining
four LEC's.  The values obtained by reproducing the experimental $np$ cross section
and magnetic moments of the deuteron and trinucleons are listed in Table~\ref{tb:tab1}.
Note that the adimensional values reported there are in units of powers of $\Lambda$, {\it i.e.}, we
have defined $d_9^{\, \prime}=d^S_1/ \Lambda^2$, $C_{15}^\prime=d^S_2/ \Lambda^4$,
$d_{21}^{\, \prime}=d^V_1/ \Lambda^2$, and $C_{16}^\prime=d^V_2/\Lambda^4$
and the superscripts $S$ and $V$ denote the isoscalar and isovector content of the
associated operators.

\begin{table}[bthp]
\vspace{-0.5cm}
\caption{Adimensional values of the LEC's corresponding to cutoff parameters $\Lambda$
in the range 500--700 MeV obtained for the AV18/UIX (N3LO/N2LO) Hamiltonian.  See text for explanation.}
\begin{tabular}{c|c|c|c|c}
\hline
  $\Lambda$   & $d^S_1\times 10^2$ & $d^S_2$ & $d^V_1$ & $d^V_2$  \\
\hline
500  & --8.85 (--0.225)  &  --3.18 (--2.38) & 5.18 (5.82) & --11.3 (--11.4) \\
600  & --2.90 (9.20)  &   --7.10 (--5.30)   & 6.55 (6.85) & --12.9 (--23.3)  \\
700  &  6.64 (20.4) & --13.2 (--9.83) &  8.24 (8.27) & --1.70 (--46.2) \\
\hline
\end{tabular}
\label{tb:tab1}
\end{table}

\begin{figure}[t]
\includegraphics[width=3in]{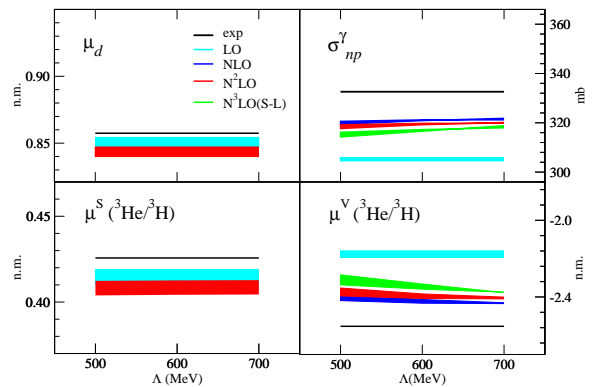}
\vspace{-0.25cm}
\caption{Results for the deuteron and trinucleon
isoscalar and isovector magnetic moments, and $np$ radiative
capture, obtained by including cumulatively the LO, NLO,
N$^2$LO, and N$^3$LO(S-L) contributions.
See text for discussion.}
\vspace{-0.5cm}
\label{fig:f3}
\end{figure}

In Fig.~\ref{fig:f3} we show results obtained by including cumulatively
the contributions at LO, NLO, N$^2$LO, and N$^3$LO(S-L) for the
deuteron ($\mu_d$) and $^3$He/$^3$H isoscalar ($\mu^S$) magnetic
moments (left panels), and for the $np$ radiative capture cross section
($\sigma_{np}^\gamma$) at thermal energies and $^3$He/$^3$H
isovector ($\mu^V$) magnetic moment (right panels).
The NLO and N$^3$LO(S-L) $M1$ operators are purely isovector, and
hence do not contribute to $\mu_d$ and $\mu^S$.  The band represents the spread in
the calculated values corresponding to the two Hamiltonian models considered
here (AV18/UIX and N3LO/N2LO).  The sensitivity to short-range
mechanisms, encoded in the cutoff $C_\Lambda(k)$ and
in the rather different short-range behaviors of the adopted potentials,
remains quite weak for all these observables.  Of course, taking into account
the N$^3$LO(LECs) contribution with the LEC values listed
in Table~\ref{tb:tab1} reproduces the experimental data
represented by the black band (to accommodate errors, but
these are negligible in the present case).
The contributions at LO and NLO have
the same sign, while those at N$^2$LO and N$^3$LO(S-L)
have each opposite sign, and tend to increase the difference
between theory and experiment.

\begin{figure}[t]
\includegraphics[width=2.7in]{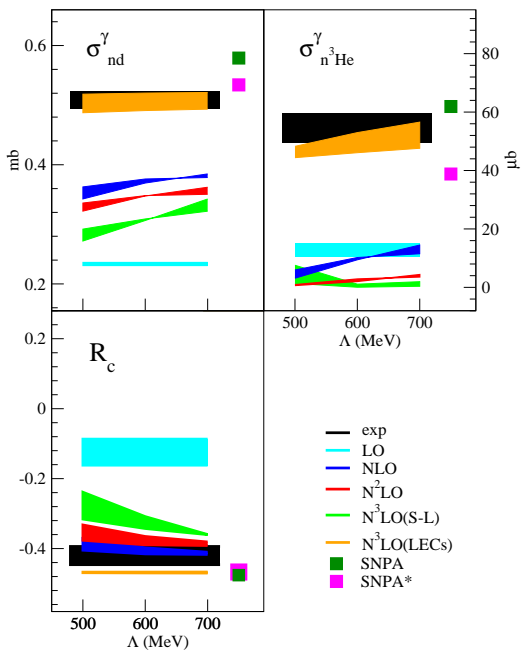}
\vspace{-0.3cm}
\caption{Results for $\sigma_{nd}^\gamma$ (left top panel),
$\sigma_{n\, ^3{\rm He}}^\gamma$ (right top panel),
and $R_c$ (left bottom panel), obtained by including cumulatively the LO, NLO,
N$^2$LO, N$^3$LO(S-L), and N$^3$LO(LECs) contributions.  Also shown are
predictions obtained in the standard nuclear physics approach (squares labeled
SNPA and SNPA$^*$).  See text for discussion.}
\vspace{-0.5cm}
\label{fig:f4}
\end{figure}

Having fully constrained the $\chi$EFT $M1$ operator up to N$^3$LO,
we are now in a position to present predictions, shown in
Fig.~\ref{fig:f4}, for the $nd$ and $n\, ^3$He radiative
capture cross sections, denoted as $\sigma_{nd}^\gamma$
and $\sigma_{n\, ^3{\rm He}}^\gamma$, and the photon
circular polarization parameter $R_c$ resulting from the
capture of polarized neutrons on deuterons.  The experimental
data (black bands) are from Ref.~\cite{Jurney82} for $nd$ and Ref.~\cite{Wolfs89} for $n\, ^3$He.
Results obtained with the complete N$^3$LO $\chi$EFT operator are shown by the orange
band labeled N$^3$LO(LECs): those corresponding to the AV18/UIX (N3LO/N2LO)
model delimit the upper (lower) end of the band in the case of $nd$, and
its lower (upper) end in the case of $n\, ^3$He.  Their sensitivity to the
cutoff, within a given model, is negligible for $nd$ and at the 5--10\% level for $n\, ^3$He.
The AV18/UIX and N3LO/N2LO results are within $\simeq 2$\% of the $nd$ experimental
cross section.  However, at $\Lambda$=600 MeV, for example, the experimental
$\sigma_{n\, ^3{\rm He}}^\gamma$ is well reproduced in the N3LO/N2LO calculation,
but underpredicted by $\simeq 15$\% in the AV18/UIX.  As expected,
these processes are strongly suppressed at LO:
the calculated $\sigma_{nd}^\gamma$(LO) and $\sigma_{n\, ^3{\rm He}}^\gamma$(LO)
are less than half and a factor of five smaller than the measured values.
In the case of $n\, ^3$He, the matrix element at NLO is
of opposite sign and twice as large (in magnitude) compared to that at LO, hence
$\sigma^\gamma_{n\, ^3{\rm He}}({\rm LO})$ and  $\sigma^\gamma_{n\, ^3{\rm He}}({\rm LO+NLO})$
are about the same, as seen in Fig.~\ref{fig:f4}.  For $nd$, however, the LO and NLO contributions
interfere constructively.  For both $nd$ and $n\, ^3$He, the N$^2$LO and N$^3$LO(S-L) corrections
exhibit the same pattern discussed in connection with Fig.~\ref{fig:f3}.  The N$^3$LO(LECs)
contributions are large, and essential for bringing theory into good agreement with experiment.

In Fig.~\ref{fig:f4} we also show results obtained in the
conventional framework, referred to as the standard nuclear
physics approach (SNPA), with the AV18/UIX Hamiltonian model.
The electromagnetic current operator includes the one-body term---the
same as the $\chi$EFT LO operator discussed earlier---as well
as two- and three-body terms, constructed from the two- and
three-nucleon potentials (AV18 and UIX, respectively) so as to
satisfy exactly current conservation with them, see Marcucci
{\it et al.} (2005)~\cite{Marcucci05}.  In the figure, the squares labeled
SNPA$^*$ represent the results obtained by retaining in addition
the relativistic corrections to the one-body current ({\it i.e.},
the $\chi$EFT N$^2$LO operator).  These corrections had been
neglected in all previous studies of these processes~\cite{Marcucci05}.
In fact, their contributions are found to be numerically significant and,
at least for the case of $nd$ capture, bring the present SNPA$^*$
results within $\simeq 4$\% and 2\%, respectively, of the experimental data and
$\chi$EFT predictions (based on the AV18/UIX model).  However,
it should be emphasized that the SNPA (and SNPA$^*$) currents contain
no free parameters---{\it i.e.}, they are not constrained to fit
any photonuclear data, in contrast to the $\chi$EFT currents.
From this perspective, the achieved level of agreement between
SNPA$^*$ and data should be viewed as satisfactory,
especially when considering, in the $\chi$EFT context,
the large role played by the N$^3$LO(LECs) currents.

We conclude by remarking that the convergence of the chiral
expansion is problematic for these processes.  The LO is
unnaturally small, since the associated
operator cannot connect the dominant S-states in the hydrogen
and helium bound states (in contrast to $np$ capture,
for example)~\cite{Marcucci05}.  This leads to an enhancement of
the NLO, which, however, in the case of $n\,^3$He is offset by
the destructive (and accidental) interference between it and the LO
contribution.  It appears that at N$^4$LO no additional LEC's enter,
Park {\it et al.} (2000)~\cite{Park00}.  Thus, inclusion
of the N$^4$LO currents would have to be followed by a ``rescaling'' of the LEC's in
Table~\ref{tb:tab1}, in order to reproduce (as before at N$^3$LO) the experimental values
of $\mu_d$, $\mu^{S,V}$, and $\sigma_{np}^\gamma$.  The  resulting predictions
for $\sigma^\gamma_{nd}$ and $\sigma^\gamma_{n\, ^3{\rm He}}$ would presumably
be close to those obtained here at N$^3$LO.

It is likely that explicit inclusion of $\Delta$ degrees of
freedom would significantly improve the convergence pattern, particularly in view
of the relevance of the pion-exchange current of panel (j) in
Fig.~\ref{fig:f1}---in such a theory, this operator would be promoted
to N$^2$LO~\cite{Pastore09}.  We plan to pursue vigorously
this line of research in the future.

The work of R.S.\ is supported by the U.S.~Department of Energy,
Office of Nuclear Physics under contract DE-AC05-06OR23177.
\end{document}